\documentclass[reprint,amsmath,amssymb,prb,]{revtex4-1}
 
\usepackage{graphicx}% Include figure files
\usepackage{dcolumn}% align table columns on decimal point
\usepackage[dvipsnames]{xcolor}
\usepackage{accents}
\usepackage{bm}
\usepackage{bbold}
\usepackage{multirow}  

\usepackage{subfiles} % Best loaded last in the preamble

\definecolor{darkgray}{rgb}{0.3, 0.3, 0.3}

\begin{document}
%\onecolumngrid

\title{ Josephson effect of superconductors with $J=3/2$ electrons}
%\altaffiliation[Also at ]{}%Lines break automatically or can be forced with \\
\author{Dakyeong Kim$^{1}$}
\author{Shingo Kobayashi$^{2}$}
\author{Yasuhiro Asano$^{1,3}$}
%\email{\empty}
\affiliation{$^{1}$ Department of Applied Physics,
Hokkaido University, Sapporo 060-8628, Japan\\
$^{2}$ RIKEN Center for Emergent Matter Science, Wako, Saitama
351-0198, Japan \\
$^{3}$ Center of Topological Science and Technology,
Hokkaido University, Sapporo 060-8628, Japan\\
}%

\date{\today}% It is always \today, today,

\begin{abstract}
The angular momentum of an electron is characterized well by 
pseudospin with $J=3/2$ in the presence of strong spin-orbit interactions.
We study theoretically the Josephson effect of superconductors 
in which such two $J=3/2$ electrons form a Cooper pair.
Within even-parity symmetry class, pseudospin-quintet pairing states with $J=2$ can exist as well as pseudospin-singlet state with $J=0$.
We focus especially on the Josephson selection rule among these even-parity superconductors.
We find that the selection rule between quintet states is severer than that 
between spin-triplet states formed by two $S=1/2$ electrons.
The effects of a pseudospin-active interface on the selection rule 
are discussed as well as those of
odd-frequency Cooper pairs generated by pseudospin dependent band structures.
\end{abstract}

%\pacs{74.81.Fa, 74.25.-q, 74.45.+c}

\maketitle

%%%%%%%%%%%%%%%%%%%%%%%%%%

%=====================================================
\section{Introduction}
%=====================================================

Spin-orbit interaction is a source of exotic electronic states realized in topological semimetals, 
topological insulators and topological superconductors.\cite{RevModPhys.82.3045,qi2011topological,tanaka2011symmetry,ando2013topological,ando2015topological,sato2016majorana,armitage2018weyl}
Recently, the ternary half Heusler compounds present another platform for systematically studying the coexistence of strong spin-orbit interaction and superconductivity.\cite{butch:prb2011,tafti:prb2013,bay:prb2012,kim2018beyond}
As a result of the strong coupling between spin $S=1/2$ and orbital angular momentum $L=1$, the total angular momentum of an electron can be $J=L+S=3/2$.
Recent studies have suggested a possibility of superconductivity due to Cooper pairing between two electrons with pseudospin of $J=3/2$.\cite{brydon2016pairing,boettcher:prb2016,savary2017superconductivity,boettcher2018unconventional}
In the case of a Cooper pair consisting of two spin $S=1/2$ electrons, it is widely accepted that spin-singlet and spin-triplet are possible pairing states for even- and odd-parity symmetry classes, respectively.
The pairing states of two $J=3/2$ electrons have richer variety.\cite{agterberg2017bogoliubov,venderbos:prx2018,kawakami:prx2018,yu2018singlet}
Namely, in addition to a pseudospin-singlet state with $J=0$, pseudospin-quintet states with $J=2$ are possible for even-parity class.
For odd-parity class, pseudospin-septet states with $J=3$ are possible as well as pseudospin-triplet states with $J=1$.
Such high angular-momentum pairing states would feature superconducting phenomena of a $J=3/2$ superconductor.
At present, however, only little has been known for physics of superconductors with $J=3/2$ electrons.\cite{ghorashi:prb2017,kobayashi:prl2019}
In this paper, we discuss a low-energy transport property unique to $J=3/2$-electron superconductors.

The Josephson effect is a fundamental property of all superconducting junctions consisting of more than one superconductor.\cite{josephson:physlett1962}
When an insulator separates two superconductors, the dissipationless electric current $I$ flows in the presence of the phase difference $\varphi$ across the junction.
The internal structure of Cooper pairs can be studied via the Josephson current because it depends sensitively on the pairing symmetries of the two superconductors.
Just below the superconducting transition temperature $T_c$, the current-phase ($I-\varphi$) relationship (CPR) can be described by
\begin{align}
  I=I_1 \sin\varphi -I_2 \sin 2\varphi, \label{Eq:CPR}
\end{align}
in all Josephson junctions consisting of time-reversal symmetry respecting materials.
The lowest order term $I_1$ disappears when the Cooper pairing states in the two superconductors 
are orthogonal to each other.
%It is a selection rule which determines whether finite lowest order term is allowed or not.
In a case of two spin-triplet superconductors of $S=1/2$ electrons, for instance, 
$I_1\propto  \boldsymbol{d}_L\cdot \boldsymbol{d}_R$ represents a selection rule for spin-triplet states, where $\boldsymbol{d}_j$ with $j=L$ ($R$) is $\boldsymbol{d}$-vector in the left (right) superconductor.
The second term $I_2$, on the other hand, is nonzero for most of the Josephson junctions.

In this paper, we will make clear the selection rules among even-parity $J=3/2$ superconducting states.
We theoretically calculate the Josephson current between two superconductors, where 
two $J=3/2$ electrons form an even-parity Cooper pair in each superconductor.
The Josephson current is calculated in terms of the anomalous Green's functions obtained 
by solving the Gor'kov equation.\cite{mahan}
The pairing state on the left(right) superconductor is represented by a vector $\bm{\eta}^{j}$ in five-dimensional pseudospin space with $j=L(R)$.
We find the relation $I_1\propto \boldsymbol{\eta}^{L}\cdot \boldsymbol{\eta}^{R}$ represents a severer selection rule of the Josephson effect in $J=3/2$-electron superconductors 
than that in $S=1/2$-electron superconductors. 
We also discuss how a magnetically active junction interface relaxes the severe selection rule 
and how subdominant pairing correlations generated by pseudospin-dependent 
dispersion affect the Josephson current.

This paper is organized as follows.
In Sec.~II, we describe a time-reversal superconducting state belonging to even-parity symmetry class for a Cooper pair formed by two $J=3/2$ electrons.
We formulate the Josephson current by using tunneling Hamiltonian between two superconductors.
In Sec.~III, the Josephson selection rule among $J=3/2$ superconducting states is studied.
In Sec.~IV, we discuss odd-frequency pairs in anisotropic band structure and following subdominant pairing correlations.
The conclusion is given in Sec.~V.
Throughout this paper, we use the units of $k_\mathrm{B}=c=\hbar=1$, 
where $k_{\mathrm{B}}$ is the Boltzmann constant and $c$ is the speed of light.

%=====================================================
\section{ A Cooper pair consisting of two $\boldsymbol{J=3/2}$ electrons }
%=====================================================

%---------------------------------------
\subsection{BdG Hamiltonian}
%---------------------------------------
Let us consider a superconductor in which two $J=3/2$ electrons form a 
Cooper pair belonging to
even-parity symmetry.
Its Hamiltonian of the normal state is given by
\begin{align}
	\mathcal{H}_N &= \sum_{\bm{k}}\Psi_{\bm{k}}^{\dagger}H_N(\bm{k})\Psi_{\bm{k}}, \\
	\Psi_{\bm{k}} &= [c_{\bm{k},3/2},c_{\bm{k},1/2},c_{\bm{k},-1/2},c_{\bm{k},-3/2}]^T ,
\end{align}
where $c_{\bm{k},j_z}$ is the annihilation operator of an electron with momentum $\bm{k}$ and the $z$-component of angular momentum being $j_z$.
The Hamiltonian with spin-orbit coupling is described as
\begin{align}
	H_N(\bm{k})&=a_1\bm{k}^2+2a_2\sum_{i}k_i^2J_i^2+a_3\sum_{i\neq j}k_ik_jJ_iJ_j - \mu, \label{Eq:normal_Ham_kp}
\end{align}
with $i=1-3$.\cite{luttinger1956quantum}
The three matrices describing angular momentum $J=3/2$ are given by
\begin{align}
	J_1 &= \frac{1}{2} \begin{pmatrix}
		0 & \sqrt{3} & 0 & 0 \\
		\sqrt{3} & 0 & 2 & 0 \\
		0 & 2 & 0 & \sqrt{3} \\
		0 & 0 & \sqrt{3} & 0
	\end{pmatrix} , \nonumber \\
	J_2 &=  \frac{i}{2}  \begin{pmatrix}
		0 &  -\sqrt{3} & 0 & 0 \\
		\sqrt{3} & 0 & -2 & 0 \\
		0 & 2 & 0 & -\sqrt{3} \\
		0 & 0 & \sqrt{3} & 0
	\end{pmatrix}, \nonumber \\
	J_3 &=  \frac{1}{2}\begin{pmatrix}
		3 & 0 & 0 & 0 \\
		0 & 1 & 0 & 0 \\
		0 & 0 & -1 & 0 \\
		0 & 0 & 0 & -3
	\end{pmatrix} .
\end{align}
The pseudospin part of a pairing state 
%in even-parity symmetry class 
is described by $\gamma$ matrices,\cite{brydon2018bogoliubov}
\begin{align}
	\gamma^0 &=1_{4 \times 4} ,\quad \gamma^1=\left(J_1\, J_2 +J_2\, J_1\right)/\sqrt{3}, \nonumber\\
	\gamma^2&=\left(J_2\, J_3+J_3\, J_2\right)/\sqrt{3} ,\quad\gamma^3=\left(J_1\, J_3+J_3\, J_1\right)/\sqrt{3} , \nonumber\\
	\gamma^4&=\left(J_1^2-J_2^2\right)/\sqrt{3} ,\quad 
	\gamma^5=\left(2J_3^2-J_1^2-J_2^2\right)/3.
\end{align}
The five matrices $\gamma^{\nu}$ for $\nu=1-5$ anticommute to one another as $\gamma^{\mu}\gamma^{\nu}+\gamma^{\nu}\gamma^{\mu}=2\gamma^0 \delta_{\mu\nu}$.
The pseudospin matrices can be expressed in terms of $\gamma^j$,
\begin{align}
J_1=& \frac{-i}{2}\left(\sqrt{3} \gamma^2\, \gamma^5 +\gamma^1\, \gamma^3 + \gamma^2\, \gamma^4 \right), \\
J_2=& \frac{i}{2}\left(\sqrt{3} \gamma^3\, \gamma^5 +\gamma^1\, \gamma^2 - \gamma^3\, \gamma^4 \right), \\
J_3=& \frac{i}{2}\left( \gamma^2\, \gamma^3 +2 \, \gamma^1\, \gamma^4 \right). 
\end{align}
The Hamiltonian in Eq.~(\ref{Eq:normal_Ham_kp}) can be transformed into %the form%~\FG{(Ex.~\ref{Ex:Cubic_Hamiltonian})}
\begin{align}
	H_N (\bm{k}) &= \varepsilon_0(\bm{k})+\sum_{\nu=1}^5 \varepsilon_{\nu}(\bm{k})\gamma^{\nu} -\mu = \mathbb{E}_{\bm{k}}, \label{Eq:normal_Ham_gamma}
\end{align}
with coefficients,
\begin{align}
	\varepsilon_0(\bm{k}) &= \frac{|\bm{k}|^2}{2m_0}, \nonumber\\
	\varepsilon_{j=1-3}(\bm{k}) &= \frac{|\bm{k}|^2}{2m_1} e_{j=1-3}(\hat{\bm{k}}) ,  \nonumber\\
	\varepsilon_{j=4,5}(\bm{k}) &=  \frac{|\bm{k}|^2}{2m_2} e_{j=4,5}(\hat{\bm{k}}),  \nonumber\\
		m_0 &= \frac{1}{2} \left(a_1+\frac{5}{4}a_2\right)^{-1}  ,\nonumber\\
	m_1 &= \frac{1}{a_3}, \quad m_2 = \frac{1}{2a_2}. \label{Eq:5_unit_vector}
\end{align}
The functions are defined as
\begin{align}	
	e_1(\hat{\bm{k}}) &= \sqrt{3}\hat{k}_x\hat{k}_y, \quad e_2(\hat{\bm{k}}) = \sqrt{3}\hat{k}_y\hat{k}_z ,\nonumber\\
	e_3(\hat{\bm{k}}) &= \sqrt{3}\hat{k}_z\hat{k}_x, \quad e_4(\hat{\bm{k}}) = \frac{\sqrt{3}}{2}(\hat{k}_x^2-\hat{k}_y^2) , \nonumber\\
	e_5(\hat{\bm{k}}) &= \frac{1}{2}(2\hat{k}_z^2-\hat{k}_x^2-\hat{k}_y^2) , \label{Eq:5_basis_function}
\end{align}
where $\hat{k}_i=k_i/|\bm{k}|$ and $e_{\nu}(\hat{\bm{k}})$ is normalized, {\it {i.e.}} $\sum_{\nu=1}^5 e_{\nu}^2(\hat{\bm{k}})=1$.
Time-reversal symmetry (TRS) of $H_N$ is represented by%~\FG{(Ex.~\ref{Ex:Symmetries_operators})}
\begin{align}
	\mathcal{T} H_N(-\bm{k})\mathcal{T}^{-1}&= H_N(\bm{k}),\\
	\mathcal{T}= U_T\mathcal{K}, \quad	U_T &= \gamma^1\gamma^2,
\end{align}
where $\mathcal{K}$ means the complex conjugation.
Under the time-reversal operation, $\gamma^{\nu}$ and $J_i$ show the relation,
\begin{align}
&\mathcal{T} \, \gamma^\nu\, \mathcal{T}^{-1} = - U_T\, (\gamma^\nu)^\ast \, U_T = \gamma^\nu,\quad \nu=0-5, \label{eq:cc_g}\\
&\mathcal{T}\,  J_i\, \mathcal{T}^{-1}= - U_T\, (J_i)^\ast \, U_T = - J_i, \quad  i = 1-3 , \label{eq:cc_j}
\end{align}
therefore $\gamma^{\nu}$ remains unchanged, whereas $J_i$ changes its sign.

A pair potential in even-parity symmetry is classified into two symmetry class in terms of its pseudospin configuration:
pseudospin-singlet and pseudospin-quintet.
Also, the symmetry class is divided to corresponding representation of crystal symmetry of the superconductor.
We list the pair potentials in a cubic symmetric superconductor with on-site~(s-wave) symmetry in Table~\ref{Table:cubic_pair_potential}.
%
%%%%%%%%%%%%%%%%%%%%%%%%%%%%%%%%%%%%%%%%%%%%%%%%%%%%%
\begin{table}
	\caption{\label{Table:cubic_pair_potential} Classification of on-site~(s-wave) symmetric Cooper pairs in a $J=3/2$ superconductor with $O_h$ symmetry, corresponding to the representation to which they belong.\cite{yang2016topological, brydon2018bogoliubov} The Cooper pair is in pseudospin-singlet($J=0$) or -quintet($J=2$) state.}
	\begin{ruledtabular}
	\begin{tabular}{ccc}
  J & Representation & Pair potential \\ \hline
  Singlet  & $A_{1g}$ &  $1_{4\times 4}$ \\
  \multirow{2}*{Quintet} & $T_{2g}$ & $  \{\gamma^1,\gamma^2,\gamma^3\}  $ \\
  ~ & $E_{g}$ & $ \{\gamma^4,\gamma^5\}$ \\
	\end{tabular}
	\end{ruledtabular}
\end{table}
%%%%%%%%%%%%%%%%%%%%%%%%%%%%%%%%%%%%%%%%%%%%%%%%%%%%%

In this paper, we focus on a superconductor preserving TRS and inversion symmetry.
The pair potential of such a superconductor
has the form\cite{brydon2018bogoliubov}
\begin{align}
	\Delta(\bm{k}) = \eta_{\bm{k},0}U_T+\bm{\eta}_{\bm{k}}\cdot\bm{\gamma} U_T . \label{eq:g-pair}
\end{align}
The first term refers to the pair potential of pseudospin-singlet state and 
the second term refers to that of pseudospin-quintet states.
The amplitudes $\eta_{\bm{k},0}$ and $\bm{\eta}_{\bm{k}}$ are even in momentum and real 
in the presence of TRS.
It is easy to confirm that Eq.~(\ref{eq:g-pair}) preserves time-reversal symmetry,
\begin{align}
\mathcal{T}\, \Delta(\boldsymbol{k})\, \mathcal{T}^{-1}= \Delta(\boldsymbol{k}),
\end{align}
due to Eq.~(\ref{eq:cc_g}).
Because electrons obey the Fermi-Dirac statistics, the pair potential must be antisymmetric under the permutation of two electrons forming a Cooper pair,
\begin{align}
\Delta^{\mathrm{T}}(-\boldsymbol{k})= -\Delta(\boldsymbol{k}),
\end{align}
where the permutation of pseudospin is carried out by taking the transpose of the matrix ($\mathrm{T}$).
The pair potentials under the consideration are even-parity, 
[i.e., $\Delta(\boldsymbol{k})=\Delta(-\boldsymbol{k})]$, 
and odd-pseudospin symmetry. 
Therefore, the relation
\begin{align}
\Delta^{\mathrm{T}}(\boldsymbol{k})= -\Delta(\boldsymbol{k}),
\end{align}
is satisfied because of 
$(\gamma^\nu)^{\mathrm{T}}=(\gamma^\nu)^\ast$ and Eq.~(\ref{eq:cc_g}).
We will use an expression $\mathbb{D}$ defined by $\Delta(\bm{k})=\mathbb{D}_{\bm{k}}U_T$, for simplicity.

The BdG Hamiltonian is expressed in terms of the normal Hamiltonian $H_N$ and the pair potential $\Delta(\bm{k})$ as% \FG{(Ex.~\ref{Ex:BdG_Hamiltonian})}
\begin{align}
	\check{H}_{\mathrm{BdG}}(\bm{k}) = \begin{bmatrix}
		H_N(\bm{k}) & \Delta(\bm{k}) \\
		-\Delta^*(-\bm{k}) & -H_N^*(-\bm{k})
	\end{bmatrix}, \label{eq:bdg}
\end{align}
which preserves particle-hole symmetry,
\begin{align}
	\mathcal{C} \check{H}_{\mathrm{BdG}}(-\bm{k})\mathcal{C}^{-1} = -\check{H}_{\mathrm{BdG}}(\bm{k}),\quad \mathcal{C}=\tau_{1}\mathcal{K},
\end{align}
where $\tau_{i}$ for $i=1-3$ is the Pauli matrix in the particle-hole space.
The Gor'kov equation becomes
\begin{align}
 \begin{bmatrix}
		i{\omega_n} -\check{H}_{\mathrm{BdG}}(\bm{k})
	\end{bmatrix}\begin{bmatrix}
		{G}(\bm{k},i{\omega_n}) & {F}(\bm{k},i{\omega_n}) \\
		\underline{{F}}(\bm{k},i{\omega_n}) & \underline{ {G}}(\bm{k},i{\omega_n})
	\end{bmatrix}=	\check{1}_{8\times 8} ,
\end{align}
where the Matsubara Green's functions are defined by
\begin{align}
G(\boldsymbol{k}, i\omega_n) =& - \int_0^{1/T} d\tau\, 
\left\langle T_\tau c_{\boldsymbol{k}, s}(\tau) c^\dagger_{\boldsymbol{k}, s^\prime}
\right \rangle\, e^{i\omega_n \tau},\\
F(\boldsymbol{k}, i\omega_n) =& - \int_0^{1/T} d\tau\, 
\left\langle T_\tau c_{\boldsymbol{k}, s}(\tau) c_{\boldsymbol{k}, s^\prime}
\right \rangle\, e^{i\omega_n \tau}.
\end{align}
Each Green's function is a $4 \times 4$ matrix.
The relations
\begin{align}
\underline{G}(\boldsymbol{k}, i\omega_n)=&- G^\ast(-\boldsymbol{k}, i\omega_n), \\
\underline{F}(\boldsymbol{k}, i\omega_n)=&- F^\ast(-\boldsymbol{k}, i\omega_n),
\end{align}
can be derived due to the particle-hole symmetry of the BdG Hamiltonian.
The solution of the BdG equation is given by
\begin{align}
&\left[\begin{array}{cc} G_(\boldsymbol{k}, i\omega_n) & F(\boldsymbol{k}, i\omega_n)\\
\underline{F}(\boldsymbol{k}, i\omega_n) &  \underline{G}(\boldsymbol{k}, i\omega_n)
\end{array}\right]
=
\left[\begin{array}{cc} 1& 0\\
0& -U_T
\end{array}\right]\nonumber\\
&\times \left[\begin{array}{cc}
i\omega_n  - \mathbb{E}_{\boldsymbol{k}} & - \mathbb{D}_{\boldsymbol{k}} \\
-  \mathbb{D}_{\boldsymbol{k}}  & i\omega_n  + \mathbb{E}_{\boldsymbol{k}}
\end{array}\right]
^{-1} 
\left[\begin{array}{cc} 1& 0\\
0& U_T
\end{array}\right] .
\end{align}
The anomalous Green's functions are calculated as
\begin{align}
 F(\boldsymbol{k}, i\omega_n) =& \mathbb{F}(\boldsymbol{k},i\omega_n)
\, U_T, \label{eq:sf}\\
 \underline{F}(\boldsymbol{k}, i\omega_n) =&  -U_T  
\mathbb{F}(\boldsymbol{k},-i\omega_n),\label{eq:sfb}\\
% \left[
% \mathbb{D} 
%  - (i\omega_n - \mathbb{E}) \mathbb{D}^{-1}
% (i\omega_n + \mathbb{E}) \right]^{-1}. 
 \mathbb{F}(\boldsymbol{k},i\omega_n) =&\left\{
 - \mathbb{D}_{\boldsymbol{k}} 
  + (i\omega_n  + \mathbb{E}_{\boldsymbol{k}}) \mathbb{D}_{\boldsymbol{k}}^{-1}
 (i\omega_n  - \mathbb{E}_{\boldsymbol{k}}) \right\}^{-1} .  \label{Eq:Gree_function}
 \end{align} 
Especially, when the dispersion relation is independent of the pseudospin, {\it {i.e.}} $\mathbb{E}_{\bm{k}}=\varepsilon_0(\bm{k})\gamma^0$, Eq.~(\ref{Eq:Gree_function}) is simplified to
 \begin{align}
  	\mathbb{F}(\bm{k},i\omega_n) &= -
	\frac{ \mathbb{D}_{\bm{k}} }{\sum_{\nu=0-5}|\eta_{\nu}|^2 + \omega_n^2 
+ \xi_{\bm{k}}^2 }, \label{Eq:ano_Green} \\
	\xi_{\bm{k}} &= \varepsilon_0(\bm{k})-\mu ,
 \end{align}
for both pseudospin-singlet or pseudospin-quintet states.

%---------------------------------------
\subsection{Current formula}
\begin{figure}[t]
	\centering
	\includegraphics[width=7.0cm]{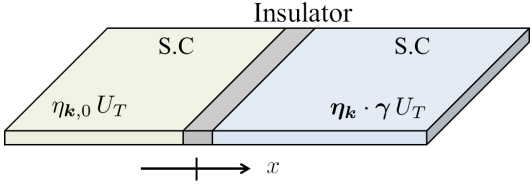}
	\caption{A schematic view of SIS Josephson junction of two $J=3/2$ superconductors and an insulator. The figure shows a junction consisting of a pseudospin-singlet and a pseudospin-quintet superconductor for instance. The insulating barrier is localized at $x=0$.}
	\label{Fig:SIQ}
\end{figure}

%---------------------------------------
We discuss the Josephson effect in a superconductor/insulator/superconductor(SIS) junction consisting of two $J=3/2$ superconductors. The Josephson current flows in the $x$ direction and 
the interface is located at $x=0$ as shown in Fig.~\ref{Fig:SIQ}.
The Hamiltonian of the SIS junction consists of three components,
\begin{align}
	H &= H_L+ H_R+ H_T,\\	
	H_T &= \sum_{\beta,\gamma}(t_{\gamma\beta}c_{\gamma}^{\dagger}c_{\beta} + t_{\beta\gamma}c_{\beta}^{\dagger}c_{\gamma}),
\end{align}
where $H_L (H_R)$ is the BdG Hamiltonian in Eq.~(\ref{eq:bdg}) of the left (right) superconductor.
$H_T$ is a tunneling Hamiltonian which refers to the interaction between two superconductors, where $\beta =\{\bm{p},s\}(\gamma = \{\bm{k},s'\})$ is a combined  index for an electron in the left(right) superconductor.
To satisfy the Hermicity of the Hamiltonian, $t_{\beta\gamma}^*=t_{\gamma\beta}$ is required.

The electric Josephson current through the junction is derived within the linear response theory with respect 
to $H_T$.\cite{mahan}
The result is expressed as%~\FG{(Ex.~\ref{Ex:linear_response_theory})}
\begin{align}
	I &= 2e \,  {\mathrm{Im}}\, T\sum_{\omega_n} \sum_{\bm{k},\bm{p}} { }^{'}   \nonumber\\
	&\times {\text {Tr}}\left[ t_{-\bm{p},-\bm{k}}^\ast\,  \underline{F}^R(\bm{k},i\omega_n) \,t_{\bm{k},\bm{p}} \, F^L(\bm{p},i\omega_n)\right], \label{Eq:Josephson_current}
\end{align}
where $\mathrm{Tr}$ means the summation over the pseudospin space, $F^{L(R)}$ is the anomalous Green's function of the left(right) superconductor and $t_{\bm{k},\bm{p} }$ represents the tunnel element, $\{ t_{\bm{k},\bm{p}} \}_{s,s'}= t_{\beta \gamma} $.
The wave number $\bm{k}$ is decomposed into a component in the $x$ direction $k_x$ and those parallel to the interface $\bm{k}_{\parallel}$. 
The summation $\sum_{\bm{k}}' $ is carried out over $\bm{k}$ for $k_x>0$.
We assume that the tunnel elements are independent of the wave numbers as
\begin{align}
	t_{\bm{k},\bm{p}}= \mathbb{T}_+\, 
	\delta_{\hat{\boldsymbol{k}}, \hat{\boldsymbol{p}}}, \quad	
 \mathbb{T}_\pm = t_0 \pm \sum_{i=1}^3 t_{i}J_i,  \label{Eq:tunneling}
\end{align}
where $\delta_{\hat{\boldsymbol{k}}, \hat{\boldsymbol{p}}}$ 
implies the presence of the translation symmetry in the direction parallel to the interface.
The tunnel event happens between two states on the Fermi surface : one is a state at $k_F \hat{\boldsymbol{p}}$ in the left superconductor and the other is a state at $k_F \hat{\boldsymbol{k}}$ in the right superconductor.
The coefficients $t_i$ are real numbers for $i=0-3$.
The element $t_0$ denotes the tunnel amplitude preserving pseudospin, whereas $t_i$ for $i=1-3$ refer to a Zeeman field which couples angular momentum of an electron.
The complex conjugation of the hopping element is represented by
\begin{align}
t_{-\boldsymbol{k}, -\boldsymbol{p}}^\ast 
= -U_T \mathbb{T}_+ \, U_T \, \delta_{\hat{\boldsymbol{k}}, \hat{\boldsymbol{p}}} 
= \mathbb{T}_- \, \delta_{\hat{\boldsymbol{k}}, \hat{\boldsymbol{p}}},
\end{align}
because of Eqs.~(\ref{eq:cc_g}) and (\ref{eq:cc_j}).
When the superconducting phase is $\varphi_L(\varphi_R)$ on the left(right) superconductor, the Josephson current in Eq.~(\ref{Eq:Josephson_current}) is calculated as
\begin{align}
I=& 2e\,  \mathrm{Im}  
{\sum_{\boldsymbol{k}, \boldsymbol{p}}}^\prime
\, e^{i\varphi} \; T  \sum_{\omega_n} \nonumber\\
&\times\mathrm{Tr}\left[ 
\mathbb{T}^{\prime}\, \mathbb{F}^R(\boldsymbol{k},-i\omega_n)\, \mathbb{T}_+\, 
\mathbb{F}^L(\boldsymbol{p},i\omega_n)  \right] \delta_{\hat{\bm{k}},\hat{\bm{p}}}, 
\label{eq:j2} \\
\mathbb{T}^{\prime} &=U_T\mathbb{T}_- U_T^{-1} = t_0+t_1 J_1-t_2 J_2+t_3 J_3.
\end{align}
We will calculate the lowest order term of Josephson current between two $J=3/2$ superconductors by using Eq.~(\ref{eq:j2}) in Sec.~\ref{sec:j1}.

%=====================================================
\section{Josephson effect}\label{sec:j1}
%=====================================================

The Josephson effect is sensitive to relative structures of pair potentials in two superconductors.
In $S=1/2$ superconductors, the Josephson current by the lowest order coupling vanishes between a spin-singlet and a spin-triplet superconductor, which is the part of the selection rule due to the spin structures of Cooper pairs.
In this section, we discuss the selection rule of the Josephson current between two $J=3/2$ superconductors.
We first consider $s$-wave superconductors with simple band structure as $\mathbb{E}_{\bm{k} }=\varepsilon_0(\bm{k}) -\mu$.
Namely, the dispersion is isotropic in the momentum space and is independent of pseudospin.
We also assume that the tunneling Hamiltonian is independent of 
pseudospin for a while, i.e. $\mathbb{T}_\pm =t_0 \gamma^0$.

%---------------------------------------
\subsection{Two pseudospin-singlet states }
%---------------------------------------

When both superconductors of a Josephson junction are in pseudospin-singlet state, the Josephson current should be represented by the Ambegaokar-Baratoff formula~\cite{ambegaokar1963tunneling}.
We will confirm this feature at the beginning of this section. 
The s-wave pair potentials are given by 
\begin{align}
	\mathbb{D}^L_{\bm{p}} = \Delta^L, \quad 
	\mathbb{D}^R_{\bm{k}} = \Delta^R ,
\end{align}
for the left and right superconductor, respectively.
The Josephson current in Eq.~(\ref{eq:j2}) becomes
\begin{align}
I_{\mathrm{SS}} =& 8e\, t_0^2 \, \sin\varphi \, T\sum_{\omega_n} \, K(i\omega_n) ,
\end{align}
while the summation over momentum is carried out as follows,
\begin{align}
K&(i\omega_n) =
{\sum_{\boldsymbol{p}} }^\prime
\frac{ \Delta^{L} }{ (\Delta^{L})^2 + \omega_n^2 + \xi_{\bm{p}}^2 } \nonumber\\
&\times \, {\sum_{\boldsymbol{k}}}^\prime
\frac{ \Delta^{R}  }{ (\Delta^{R})^2 + \omega_n^2 + \xi_{\bm{k}}^2 }
\;  \delta_{\hat{\boldsymbol{k}}, \hat{\boldsymbol{p}}},\\
&= \left\langle
 \int \frac{d\xi_{k} \,N_0}{ (\Delta^{L})^2 + \omega_n^2 
+ \xi_{k}^2 }
 \int \frac{d\xi_{k}\, N_0 }{ (\Delta^{R})^2 + \omega_n^2 
+ \xi_{k}^2 } \right.\nonumber\\
&\times
	\left.  \Delta^L \Delta^R\right\rangle_{\hat{\bm{k}}} , 
\label{Eq:summation}\\
&=  \frac{\pi N_0}{\sqrt{\omega_n^2 + (\Delta^{L})^2}} \frac{\pi N_0}{\sqrt{\omega_n^2 + (\Delta^{R})^2}} \times \Delta^L  \Delta^R .
	\label{eq:l4}
\end{align}
The summation over the momentum is decomposed into two parts. 
One part is the summation over its amplitude $|\boldsymbol{k}|$, which can be replaced by the integration over $\xi_{\boldsymbol{k}}$. 
The density of states at the Fermi level is denoted by $N_0$.
The details of the integral is summarized in Appendix.~\ref{ap:table}.
The other part is summation over the direction of $\boldsymbol{k}$ on the Fermi surface for $k_x>0$.
It is calculated as the average over the Fermi surface,
\begin{align}
\langle X \rangle_{\hat{\bm{k}}} = \frac{1}{\pi} \int_0^{\pi} d\theta \sin^2\theta \int_{-\pi/2}^{\pi/2} d\phi \cos\phi \; X(\hat{\bm{k}}),
\end{align}
where
$k_x = k_F \sin\theta\cos\phi >0 $, $k_y=k_F \sin\theta \sin\phi$ and $k_z= k_F\cos\theta$.
The Fermi momentum is obtained from $\varepsilon_0(k_F)-\mu=0$.
Finally the Josephson current is calculated as
\begin{align}
I_{\mathrm{SS}}=& I_0 \sin\varphi,\\
I_0=&\frac{\pi}{eR_N} \, T \sum_{\omega_n}\left( \frac{\Delta^L}{\sqrt{\omega_n^2 + (\Delta^L)^2}}\right)
\left( \frac{\Delta^R}{\sqrt{\omega_n^2 + (\Delta^R)^2}}\right),\\
R_N^{-1} =& 4\times 2\pi e^2 \left(N_0 t_0\right)^2,
\end{align}
where $R_{N}$ is the normal resistance of the junction. 
As the most simple case, we assume that the amplitudes of the two pair potentials are identical to each other, $\Delta^L=\Delta^R=\Delta$.
Then the Josephson current results in
\begin{align}
I_{\mathrm{SS}}=& \frac{\pi \Delta}{2 e R_N}\, \tanh\left( \frac{\Delta}{2T} \right) 
\, \sin\varphi.
\end{align}
The expression of the current is identical to the Ambegaokar-Baratoff's formula~\cite{ambegaokar1963tunneling}.

%---------------------------------------
\subsection{ Singlet-quintet states }
%---------------------------------------
Secondly, we discuss a junction where a pseudospin-singlet $s$-wave superconductor is on the left and a pseudospin-quintet superconductor is on the right side.
The pair potentials are given by
\begin{align}
	\mathbb{D}^L_{\bm{p}} = \Delta^L , 
	\quad \mathbb{D}^R_{\bm{k}} =\Delta^R  \bm{q}\cdot \bm{\gamma}.
\end{align}
The vector $\bm{q}$ is normalized vectors in five-dimensional pseudospin space, {\it{i.e.}} $\sum_{\nu=1-5}|q_{\nu}|^2=1$.
When the tunneling elements are independent of pseudospin $\mathbb{T}_{\pm}=t_0\gamma^0$,
the lowest order of the Josephson current vanishes by the selection rule due to the pseudospin configuration 
of Cooper pairs. Pseudospin of a Cooper pair is $J=0$ in a singlet superconductor, whereas it is $J=2$ in a quintet
superconductor.
To make the Josephson current finite, an electron must change its angular momentum while tunneling the insulating barrier because of the difference in the angular momentum, $\delta J =2$.

We switch on a weak Zeeman field to make the interface magnetically active.
Then the tunnel elements become pseudospin dependent and can have finite $t_i(\ll t_0)$ for $i=1-3$ in Eq.~(\ref{Eq:tunneling}).
The Josephson current is calculated as
\begin{align}
	I_{\mathrm{SQ}} &=\frac{2\sqrt{3}t_1t_3q_3 +\sqrt{3}(t_1^2+t_2^2)q_4-(t_1^2-t_2^2-2t_3^2)q_5 }{2t_0^2}  \nonumber \\
	 &\quad \times   I_0\sin\varphi .\label{eq:isq}
\end{align}
Although the lowest order current can have finite value because of the Zeeman field, the contributions of $\nu=1,2$ terms vanish.
As far as we study, the current-phase relationship is sinusoidal even if a Zeeman field breaks time-reversal symmetry. 
The results in Eq.~(\ref{eq:isq}) imply a possibility of $\pi$-junction caused by a magnetically active interface. 
Also, we find the second order terms in the coefficient $t_i$ of the Josephson current as a feature of $J=3/2$ superconductors.
The second order implies that a magnetic field affects a Cooper pair twice during the tunneling.
The reason can be explained by considering two facts that the difference in angular momentum of Cooper 
pairs is $\delta J=2$ and that $J_i$ matrices in the tunneling Hamiltonian can change the angular 
momentum only by $\delta J=1$ for each operation.
Therefore, a Zeeman field affects a Cooper pair twice during the tunneling to compensate the difference in 
the angular momenta of Cooper pairs.
In contrast, the Josephson current between  $S=1/2$ spin-singlet and triplet superconductor is calculated as $I \propto \bm{d}\cdot\bm{U}_M$, where $\bm{U}_M$ is the magnetic scattering potential~(Appendix.~\ref{ap:SIT}).
Thus the current is written in the first order terms in the coefficients of the spin-dependent tunneling Hamiltonian.
This implies that a single operation of a Zeeman field can couple the Cooper pairs with difference of the angular momentum, $\delta S =1$.

%---------------------------------------
\subsection{Two pseudospin-quintet states }
%---------------------------------------
In this subsection, we discuss the Josephson effect in a junction consisting of two superconductors in 
pseudospin-quintet states.
First, we consider a junction of two $s$-wave pseudospin-quintet superconductors described by
\begin{align}
	\mathbb{D}^L_{\bm{p}} = \Delta^L \bm{q}^L\cdot \bm{\gamma}, \quad
	\mathbb{D}^R_{\bm{k}}  = \Delta^R \bm{q}^R\cdot \bm{\gamma} .
\label{eq:delta_quint}
\end{align}
The anomalous Green's function can be calculated as Eq.~(\ref{Eq:ano_Green}).
By substituting the anomalous Green's function in Eq.~(\ref{Eq:ano_Green}) to Eq.~(\ref{eq:j2}), the Josephson current is calculated as
\begin{align}
	I_{\mathrm{QQ}} &=  I_0 \sin\varphi \left(\bm{q}^R \cdot \bm{q}^L\right). \label{Eq:Josephson_current_QQ}
\end{align}
The result indicates that the Josephson current is given by the scalar product of two vectors, $I_{\mathrm{QQ}} \propto \boldsymbol{q}^L \cdot \boldsymbol{q}^R $, which describes the selection rule due to the pseudospin configuration of Cooper pairs.
It is easy to confirm that Josephson current cannot flow between two superconductors with Cooper pairs corresponding to different representations in Table~\ref{Table:cubic_pair_potential}.
This selection rule is similar to that in two spin-triplet superconductors of $S=1/2$ electrons,
\begin{align}
	I_{\mathrm{TT}} =&  I_1
 \frac{\Big\langle \boldsymbol{d}^L \cdot \boldsymbol{d}^R \Big\rangle_{\hat{\bm{k}}}}
 {d_l\, d_r} \, \sin\varphi, 
 \label{Eq:Josephson_current_TT} \\
 d_{l(r)}=&\sqrt{\left\langle
 \left\{\boldsymbol{d}^{L(R)}(\hat{\boldsymbol{k}})\right\}^2
 \right\rangle_{\hat{\bm{k}}} }
\end{align}
where $\boldsymbol{d}$ is the $\boldsymbol{d}$-vector in three-dimensional spin space of the spin-triplet Cooper pair with pair potential $\Delta(\hat{\boldsymbol{k}}) = i \bm{d}(\hat{\boldsymbol{k}}) \cdot \bm{\sigma} \sigma_2$.
By comparing Eq.~(\ref{Eq:Josephson_current_QQ}) and (\ref{Eq:Josephson_current_TT}), we conclude that the selection rule for quintet states is severer than that for triplet states.
This is because two vectors can be orthogonal to each other more easily in higher dimension.

Additionally, we study a case with two $J=3/2$ superconductors in which pair potentials are anisotropic in the momentum space.
With the general even-parity pair potential
\begin{align}
	\mathbb{D}^L_{\bm{p}} =   \bm{\eta}_{\bm{p}}^L \cdot \bm{\gamma} , \quad
	\mathbb{D}^R_{\bm{k}}  = \bm{\eta}_{\bm{k}}^R \cdot \bm{\gamma},
%\label{eq:delta_quint}
\end{align}
the Josephson current is calculated as
\begin{align}
	I_{\mathrm{QQ}} &= I_1 \frac{\left\langle \bm{\eta}_{\bm{k}}^R\cdot \bm{\eta}_{\bm{k}}^L \right\rangle_{\hat{\bm{k}}}}{\eta^{R}\eta^{L}}  \sin\varphi , \label{Eq:Jc_QQ_anisotropic} \\
	I_1 &= \frac{\pi}{eR_N} T \sum_{\omega_n}\frac{\eta^R}{\sqrt{\omega_n^2+(\eta^R)^2}} \frac{\eta^R}{\sqrt{\omega_n^2+(\eta^R)^2}} ,\\
	 \eta^{L(R)} &= \sqrt{\left\langle
 \left(\boldsymbol{\eta}^{L(R)}_{\boldsymbol{k}}\right)^2
 \right\rangle_{\hat{\bm{k}}} }.
\end{align}
The average of $\left\langle \bm{\eta}_{\bm{k}}^R\cdot \bm{\eta}_{\bm{k}}^L \right\rangle_{\hat{\bm{k}}}$ 
on the Fermi surface leads to additional 
selection rule for the pair potentials depending on momenta.
For example in a case of $\bm{\eta}_{\bm{k}}^L \propto e_1(\hat{\boldsymbol{k}})$ 
and $\bm{\eta}_{\bm{k}}^R \propto e_2(\hat{\boldsymbol{k}})$, the Josephson current vanishes 
irrespective of relative pseudospin configurations of Cooper pairs in the two superconductors 
because of the selection rule in momentum space 
$\left\langle e_1(\hat{\boldsymbol{k}}) e_2(\hat{\boldsymbol{k}}) \right\rangle_{\hat{\bm{k}}}=0$.

In this section, we have assumed that electronic structures are independent of pseudospin, 
i.e., $\mathbb{E}_{\bm{k} }=\varepsilon_0(\bm{k}) -\mu$.
The effects of a pseudospin-dependent electronic structures on the selection rule 
will be clarified in the next section.

%=====================================================
\section{Anisotropic Dispersion and Odd-frequency Pair}
%=====================================================

In this section, we show that anisotropic dispersion depending on pseudospin 
generates an odd-frequency Cooper pair and discuss how it affects the Josephson current. 
For simplicity, we consider a simple superconductor in which the dispersion and the pair potential are described by
\begin{align}
	\mathbb{E}_{\boldsymbol{k}} = \varepsilon_0(\bm{k})
 + \varepsilon_{\lambda}(\bm{k})\, \gamma^{\lambda} -\mu, 
	\quad \mathbb{D}_{\boldsymbol{k}}
 = \eta_{\bm{k}}\, \gamma^{\nu}. \label{Eq:anisotropic_left}
\end{align}
Here we assume $|\varepsilon_{\lambda} |\ll |\varepsilon_0|$.
When $\lambda =\nu$, $\mathbb{E}_{\bm{k}}$ and $\mathbb{D}_{\bm{k}}$ commute to each other. 
Then the symmetry of the pair potential and that of the anomalous Green's function are the same because $\mathbb{F}(\bm{k},i\omega_n)$ in Eq.~(\ref{Eq:Gree_function}) satisfies $\mathbb{F} \propto \gamma^\nu$ near the Fermi surface.
In this case, the selection rule in Eq.~(\ref{Eq:Jc_QQ_anisotropic}) remains valid. 
In other words, the selection rule should be modified when $\mathbb{E}_{\boldsymbol{k}}$ and $\mathbb{D}_{\boldsymbol{k}}$ do not commute in each superconductor, {\it{i.e.}} $\lambda \neq \nu$.
The anomalous Green's function is calculated as% \FG{(Ex.~\ref{Ex:Anisotropic_dispersion})}
\begin{align}
	\mathbb{F}_{\nu\lambda}(\bm{k},i\omega_n) &= \mathbb{F}_{\nu}^{(0)}(\bm{k},i\omega_n) + \delta \mathbb{F}_{\nu\lambda}(\bm{k},i\omega_n),  \label{Eq:perturbated_dispersion} \\
	\mathbb{F}_{\nu}^{(0)}(\bm{k},i\omega_n) &= 
	-\frac{ 1}
	{\xi_{\bm{k}}^2 + \omega_n^2 +\eta_{\bm{k}} ^2 }\eta_{\bm{k}} \gamma^{\nu}, \label{eq:f0e}\\
	\delta \mathbb{F}_{\nu\lambda}(\bm{k},i\omega_n) &= - 2i\omega_n 
	\frac{\varepsilon_{\lambda}(\boldsymbol{k})}{(\xi_{\bm{k}}^2+\omega_n^2+ \eta_{\bm{k}}^2 )^2 } \eta_{\bm{k}} \gamma^{\nu}\gamma^{\lambda}.\label{eq:f1o}
\end{align}
The details of the derivation are shown in Appendix~\ref{ap:odd}. 
The extra term $\delta \mathbb{F}_{\nu\lambda}$ describes an unusual Cooper pairing correlation belonging to odd-frequency symmetry.
Since both $\varepsilon_{\lambda}(\bm{k})$ and $\eta_{\bm{k}}$ are even-parity functions, 
$\delta F_{\nu\lambda}(\bm{k},i\omega_n)=\delta \mathbb{F}_{\nu \lambda}(\bm{k},i\omega_n)\, U_T$ is also an even-parity function.
The product $ \gamma^{\nu} \gamma^{\lambda}U_T$ suggests that $\delta F_{\nu\lambda}$ is symmetric under the permutation of two pseudospins.
To meet the requirement of the Fermi-Dirac statistics of electrons, $\delta F_{\nu\lambda}$ must be odd under the permutation of imaginary times.
As a result, $\delta F_{\nu\lambda}$ is confined to be an odd function of the Matsubara frequency.\cite{berezinskii:jetplett1974}

Mathematically speaking, the kinetic Hamiltonian depending on pseudospin, $V=\varepsilon_\lambda (\bm{k})\gamma^\lambda$ in Eq.~(\ref{Eq:anisotropic_left}), generates an odd-frequency pairing correlation because it does not commute to the pair potential.
The existence of an odd-frequency pair in an uniform superconductor was pointed out in a multiband superconductor, where the band hybridization generates an odd-frequency Cooper pair.\cite{BSchaffer:prb2013}
It has been also well established that such an odd-frequency pair is paramagnetic.\cite{asano:prb2015,asano:prl2011,suzuki:prb2014} 
A usual (even-frequency) Cooper pair is diamagnetic and excludes magnetic fields. 
Therefore, a superconductor can maintain its macroscopic phase being uniform in real space and can decrease the condensation energy. 
A paramagnetic Cooper pair, on the other hand, is unstable thermodynamically because it causes spatial variations in the superconducting phase.\cite{fominov:prb2015,mironov:prl2012,suzuki:prb2014} 
As a consequence, the appearance of odd-frequency pairing correlations decreases the superconducting transition temperature $T_c$.\cite{asano:prb2015, triola:physik2020}
\citeauthor{ramires:prb2016}\cite{ramires:prb2016} showed that a perturbation $V$ in Hamiltonian decreases the transition temperature of a superconducting order parameter $\Delta$ when 
 \begin{align}
 V(\boldsymbol{k})\, \Delta(\boldsymbol{k}) -  \Delta(\boldsymbol{k}) \,
 V^\ast(-\boldsymbol{k})
 \neq 0. \label{eq:fitness}
 \end{align}
It is easy to confirm that Eq.~(\ref{eq:fitness}) corresponds to the condition for the appearance of an odd-frequency pair, $\lambda \neq \nu$.

%-----------------------------------------
\subsection{Two identical superconductors}
%-----------------------------------------

We first consider a junction that consists of two identical superconductors described by Eq.~(\ref{Eq:anisotropic_left}) and a barrier with tunnel elements $t_{\bm{k},\bm{p}}= t_0 \gamma^0 \delta_{\hat{\boldsymbol{k}}, \hat{\boldsymbol{p}}}$.
By substituting the Green's function in Eq.~(\ref{Eq:perturbated_dispersion}) into Eq.~(\ref{eq:j2}), the Josephson current is calculated as
\begin{align}
	I
	=&2e\, t_0^2\, \sin\varphi\, T\sum_{\omega_n}{\sum_{\boldsymbol{k},\boldsymbol{p}}}^\prime\,
\mathrm{Tr}
\left[ \mathbb{F}^{R(0)}_\nu(\boldsymbol{k}, -i\omega_n) \, 
\mathbb{F}^{L(0)}_\nu(\boldsymbol{p}, i\omega_n)\right.\nonumber\\
&+\left.
\delta \mathbb{F}_{\nu\lambda}^{R}(\boldsymbol{k},  -i\omega_n)  \,
\delta \mathbb{F}_{\nu\lambda}^{L}(\boldsymbol{p}, i\omega_n)
\right],\label{eq:jodd_def}\\
	=& I_0  \sin \varphi 
	\left(
1- \frac{\langle \varepsilon_\lambda(\bm{k}) \rangle_{\hat{\boldsymbol{k}}}} {8\eta^2
}
\right).\label{eq:jodd1}
\end{align}
The details of the derivation are shown in Appendix~\ref{ap:odd}. 
 The first term of Eqs.~(\ref{eq:jodd_def})-(\ref{eq:jodd1}) has been already explained
in Sec.~\ref{sec:j1} C and it dominates the Josephson current.
An induced odd-frequency pair carries the Josephson current at the second term in Eqs.~(\ref{eq:jodd_def})-(\ref{eq:jodd1}). 
The contribution of an odd-frequency pair is negative, which is 
derived from the order of pseudospin matrices in the second term as
$\mathrm{Tr}[\gamma^\nu\, \gamma^\lambda\, \gamma^\nu\, \gamma^\lambda]=-4$.
Namely, the coupling between the induced odd-frequency pairing correlations
stabilizes the $\pi$-state rather than the 0-state
because an odd-frequency pair favors phase difference across the junction 
due to its paramagnetic property.\cite{sasaki:prb2020}
When the amplitude of odd-frequency correlation in Eq.~(\ref{eq:f1o}) 
exceeds that of even-frequency component in Eq.~(\ref{eq:f0e}), 
the junction may undergo the transition to the $\pi$-state.
However, such a superconducting state is impossible because 
an odd-frequency Cooper pair is thermodynamically unstable.

%-----------------------------------------
\subsection{Two different superconductors}
%-----------------------------------------

As shown in Eq.~(\ref{eq:jodd1}), the junction we consider should be always the 0-state because the even-frequency pairing correlations dominate the Josephson current.
Simultaneously, it proposes a possibility of a $\pi$-junction in the system consisting of time-reversal symmetric components only, if the dominant term could be deleted.
In such a context, we consider a Josephson junction where the severer pseudospin selection rule is applied.
We choose its left superconductor described by Eq.~(\ref{Eq:anisotropic_left}) and the right superconductor described by
\begin{align}
	\mathbb{E}_{\bm{k}}= \varepsilon_0 + \varepsilon_{\nu}(\bm{k})\gamma^{\nu} -\mu, 
	\quad \mathbb{D}_{\bm{k}} = \eta_{\bm{k}}^R \,\gamma^{\lambda}. \label{Eq:anisotropic_right}
\end{align}
The anomalous Green's function of the right superconductor is described by
\begin{align}
	\mathbb{F}_{\lambda\nu}(\bm{k},i\omega_n) = \mathbb{F}_{\lambda}^{(0)}(\bm{k},i\omega_n) + \delta \mathbb{F}_{\lambda\nu} (\bm{k},i\omega_n).
\end{align}
The Josephson current in this junction is calculated as
\begin{align}
I=&2e\, t_0^2\, \sin\varphi\, T\sum_{\omega_n}{\sum_{\boldsymbol{k},\bm{p}}}^\prime \,
\mathrm{Tr}\left[\mathbb{F}^{L(0)}_\nu(\boldsymbol{p}, i\omega_n) \, 
\mathbb{F}^{R(0)}_\lambda(\boldsymbol{k}, -i\omega_n) \right.\nonumber\\
&+\left.
\delta \mathbb{F}_{\nu\lambda}^L(\boldsymbol{p}, i\omega_n) \,
\delta \mathbb{F}_{\lambda\nu}^R(\boldsymbol{k}, -i\omega_n) 
\right],\label{eq:jodd2e}\\
=&\frac{\pi }{eR_N}\langle \varepsilon_\lambda(\bm{k}) \, \varepsilon_\nu(\bm{k}) 
\eta_{\bm{k}}^R\eta_{\bm{k}}^L \rangle_{\hat{\boldsymbol{k}}} 
\, \sin\varphi \nonumber\\
&\times T\sum_{\omega_n}  
\frac{ \omega_n^2 } 
{\{\omega_n^2+(\eta^R)^2\}^{3/2}\{\omega_n^2+(\eta^L)^2\}^{3/2}}.
\end{align}
The positive sign of the trace is derived from the order of pseudospin matrices as 
 $\mathrm{Tr}[\gamma^\nu\, \gamma^\lambda\, \gamma^\lambda\, \gamma^\nu]=4$.
The first term in Eq.~(\ref{eq:jodd2e}) vanishes due to the pseudospin selection rule as expected.
The second term is carried by induced odd-frequency and its sign is determined by the momentum dependence of the pair potentials.
However, it is calculated as positive or zero if the momentum dependence is $s$-wave or $d$-wave symmetric.
Unfortunately, we cannot draw any physical picture of why an odd-frequency pair stabilizes the 0-state. 
Even so, the result of the 0-junction is reasonable in the phenomenological point of view.
Because all the parts of the Josephson junction preserve time-reversal symmetry as we have discussed at the beginning of this subsection.
The superconducting phase should be uniform at the ground state of such a junction.

%=====================================================
\section{Conclusion}
%=====================================================
In this paper, we discussed the Josephson effect 
of a superconductor in which an electron with angular momentum $J=3/2$ 
characterizes the electronic structures near the Fermi level as a result of 
strong spin-orbit interactions.
We assume that a $J=3/2$ superconductor preserves time-reversal and inversion symmetries 
simultaneously. 
In the presence of inversion symmetry, $J=0$ pseudospin-singlet and $J=2$ pseudospin-quintet 
state are possible.
The latter is described by the combination of five even-parity orbital functions and 
five pseudospin tensors.
The Josephson current within the tunnel Hamiltonian description 
is calculated in terms of the anomalous Green's functions on either sides of a junction.
We found that the Josephson selection rule for quintet states is stricter than 
that for well-established $S=1$ spin-triplet states.
A magnetically active junction interface enables the Josephson coupling between a pseudospin-singlet $J=0$ superconductor and a pseudospin-quintet $J=2$ superconductor. 
We also discussed a Josephson current carried by odd-frequency pairing correlations which are generated 
by the complex commutation relations among pseudospin tensors. 
We find that the odd-frequency pairing correlation favors the $\pi$-state as long as 
it is a subdominant pairing correlation in a superconductor. 

In a $J=3/2$ superconductor, in addition to pseudospin quintet states, pseudospin-septet states with angular momentum $J=3$ can exist where a Cooper pair belongs to odd-parity symmetry.
The physics of such a high angular momentum pair and specification of the pair potential 
are issues in the future. 
The physics of topological surface states could be an another open issue.

%---------------------------------------
\begin{acknowledgments}
This work was supported by JSPS KAKENHI
(No.~JP20H01857), JSPS Core-to-Core Program (A.
Advanced Research Networks), and JSPS and Russian
Foundation for Basic Research under Japan-Russia
Research Cooperative Program Grant No. 19-52-50026.
S.K. was supported by JSPS KAKENHI Grants No. JP19K14612 and by the
CREST project (JPMJCR16F2, JPMJCR19T2) from Japan Science and Technology
Agency (JST).
\end{acknowledgments}
%---------------------------------------

%***************************************************************
\appendix
%***************************************************************

\begin{widetext}
%=======================================================================
\section{Table for integrals, summations and constants}\label{ap:table}
%=======================================================================
Here we summarize formulas of integral and summation over the Matsubara frequency, used in this paper.
\begin{align}
\int_{-\infty}^\infty\frac{dx}{x^2+a^2}=& \frac{\pi}{a},\quad
\int_{-\infty}^\infty\frac{dx}{(x^2+a^2)^2}= \frac{\pi}{2a^3},\\
T\sum_{\omega_n}\frac{1}{\omega_n^2+a^2}=&
\frac{1}{2a}\tanh\left(\frac{a}{2T}\right),\\
T\sum_{\omega_n}\frac{\omega_n^2}{(\omega_n^2+a^2)^3}=&
\frac{1}{16 a^3}\left[
\tanh\left(\frac{a}{2T}\right)
-\frac{a}{2T} \cosh^{-2}\left(\frac{a}{2T}\right)
+2 \left( \frac{a}{2T} \right)^2
\frac{\tanh\left(\frac{a}{2T}\right)}{\cosh^2\left(\frac{a}{2T}\right)}\right], \\
\approx & \frac{1}{16 a^3}\tanh\left(\frac{a}{2T}\right).
\end{align}
The last equation is approximated in case $T \ll a$.
The average of functions in Eq.~(\ref{Eq:5_basis_function}) over the Fermi surface give constants,
\begin{align}
&\left\langle e_1^2(\hat{\boldsymbol{k}}) \right\rangle_{\hat{\bm{k}}}=\frac{3}{10}, 
\quad
\left\langle e_2^2(\hat{\boldsymbol{k}}) \right\rangle_{\hat{\bm{k}}}=\frac{1}{8}, \quad
\left\langle e_3^2(\hat{\boldsymbol{k}}) \right\rangle_{\hat{\bm{k}}}=\frac{1}{4}, \quad
 \left\langle e_4^2(\hat{\boldsymbol{k}}) \right\rangle_{\hat{\bm{k}}}=\frac{7}{32},\quad
\left\langle e_5^2(\hat{\boldsymbol{k}}) \right\rangle_{\hat{\bm{k}}}=\frac{5}{32},\\
&\left\langle e_1(\hat{\boldsymbol{k}})\, e_{\lambda\neq 1}(\hat{\boldsymbol{k}}) \right\rangle_{\hat{\bm{k}}}
=\left\langle e_2(\hat{\boldsymbol{k}})\, e_{\lambda\neq 2}(\hat{\boldsymbol{k}}) \right\rangle_{\hat{\bm{k}}}
=\left\langle e_3(\hat{\boldsymbol{k}})\, e_{\lambda\neq 3}(\hat{\boldsymbol{k}}) \right\rangle_{\hat{\bm{k}}}=0, \quad \left\langle e_4(\hat{\boldsymbol{k}})\, e_5(\hat{\boldsymbol{k}}) \right\rangle_{\hat{\bm{k}}}
=\frac{\sqrt{3}}{32}.
\end{align}

%=======================================================================
\section{The Josephson effect in a spin-singlet/spin-triplet junction}\label{ap:SIT}
%=======================================================================
We summarize the selection rule of the Josephson current between two $S=1/2$ superconductors: one is in a 
spin-singlet state and the other is in a spin-triplet state on the basis of Ref.~\onlinecite{brydon2013charge}.
This gives a fine reference to study the selection rule of two $J=3/2$ superconductors.
The Josephson junction consists of spin-triplet(singlet) superconductors on the left(right) side and a magnetically active insulator at $z=0$~(Fig.~\ref{Fig:TFS}).
\begin{figure}[b]
	\includegraphics[width=7.0cm]{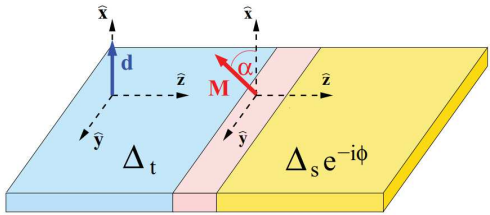}
	\caption{A schematic view of a Josephson junction consisting of a spin-triplet and singlet superconductor and an insulator.\cite{brydon2013charge} The insulating barrier is localized at $z=0$. For simplicity, the $x$-direction is set to be parallel to the $\bm{d}$-vector of the spin-triplet state.}
	\label{Fig:TFS}
\end{figure}
The pair potentials are given by $\Delta^L(\bm{k}) = i\bm{d}_{\bm{k}}\cdot\hat{\bm{\sigma}}\hat{\sigma}_2$ and $\Delta^R(\bm{k}) = i\Delta_{s,\bm{k}}\hat{\sigma}_2 e^{-i\phi}$, respectively.
$\hat{\sigma}_i$ is the Pauli matrices in spin space and $\bm{d}_{\bm{k}}$ is the $\bm{d}$-vector of the spin-triplet state.
For the barrier potential, $\hat{V}(\bm{r})=\left(U_0\hat{\sigma}_0+\bm{U}_M\cdot \hat{\bm{\sigma}}\right)\delta(z)$ is given where $U_0$ is the charge scattering potential and $\bm{U}_M=U_M(\cos\alpha\hat{\bm{e}}_x+\sin\alpha\hat{\bm{e}}_y)$ is the magnetic scattering potential. $\alpha$ is the angle between the $\bm{d}$-vector and the magnetization $\bm{M}$ in the insulator.
Using the tunneling perturbation theory, \citeauthor{brydon2013charge} calculated the Josephson current $I$ and showed
\begin{align}
	I \propto  \cos\alpha\cos\phi. \label{Eq:SIT_current}
\end{align}
The lowest order term is proportional to $\cos\phi$ because the magnetic field breaks time-reversal symmetry.
From $\cos\alpha = (\bm{U}_M\cdot \bm{d})/(|\bm{U}_M||\bm{d}|)$, we find that the current is proportional to the scalar product of $\bm{d}$-vector and $\bm{U}_M$.

%=======================================================================
\section{Odd-frequency pairing}\label{ap:odd}
%=======================================================================
Here we first derive the anomalous Green's function in a superconductors in the presence of pseudospin-dependent 
dispersion. Then we show the derivation of the Josephson current.
When a superconductor is described by Eq.~(\ref{Eq:anisotropic_left}), its anomalous Green's function is calculated as
\begin{align}
	\mathbb{F}_{\nu\lambda}(\bm{k},i\omega_n) 	 &= \left[ - \eta_{\bm{k}} \gamma^{\nu} + \left(i\omega_n +\xi_{\bm{k}} +\varepsilon_{\lambda}(\bm{k})\, \gamma^{\lambda} \right)
	\left(  \eta_{\bm{k}} \gamma^{\nu} \right)^{-1} \left( i\omega_n -\xi_{\bm{k}} -\varepsilon_{\lambda}(\bm{k})\, \gamma^{\lambda} \right)
	\right]^{-1}, \nonumber\\
	&= - \left[ \left\{  \frac{ \eta_{\bm{k}}^2 + \omega_n^2+ \xi_{\bm{k}}^2 - \varepsilon_\lambda^2(\boldsymbol{k})}{\eta_{\bm{k}} }  \gamma^{0}  - 2 i\omega_n \frac{  \varepsilon_{\lambda}(\bm{k})}	{ \eta_{\bm{k}} }  \gamma^{\lambda}\right\} \gamma^{\nu}   \right]^{-1},\nonumber \\
	&\approx  -   \frac{ \eta_{\bm{k}}}{ \eta_{\bm{k}}^2+\omega_n^2+\xi_{\bm{k}}^2 }\gamma^{\nu} -2i\omega_n \frac{\varepsilon_{\lambda}(\bm{k})  \eta_{\bm{k}} }{ ( \eta_{\bm{k}}^2+\omega_n^2+\xi_{\bm{k}}^2 )^2}\gamma^{\nu} \gamma^{\lambda}	. \label{eq:fnumu}
\end{align}
At the last line, we assumed $|\varepsilon_\lambda(\bm{k})| \ll |\varepsilon_0(\bm{k})|$ for $\lambda=1-5$.
The second term refers an odd-frequency pairing.

We construct a Josephson junction with the left superconductor described by Eq.~(\ref{Eq:anisotropic_left}) and the right superconductor described by
\begin{align}
	\mathbb{E}_{\boldsymbol{k}} = \varepsilon_0(\bm{k})
 + \varepsilon_{\rho}(\bm{k})\, \gamma^{\rho} -\mu, 
	\quad \mathbb{D}_{\boldsymbol{k}}
 = \eta_{\bm{k}}\, \gamma^{\kappa}, \label{Eq:anisotropic_other}
\end{align}
with $\rho\neq \kappa$. 
When an insulator is magnetically inactive {\it{i.e.}} $\mathbb{T} = t_0\gamma^0$, the Josephson current becomes
\begin{align}
	I =& 2e\, t_0^2\, \sin\varphi\, T {\sum_{\boldsymbol{k},\boldsymbol{p}}}^\prime \,\mathrm{Tr}\left[\mathbb{F}^0_\nu(\boldsymbol{k}, -i\omega_n) \,\mathbb{F}^0_\nu(\boldsymbol{p}, i\omega_n) +
\delta \mathbb{F}_{\nu\lambda}(\boldsymbol{k}, -i\omega_n) \,
\delta \mathbb{F}_{\nu\lambda}(\boldsymbol{p}, i\omega_n) 
\right] \delta_{\hat{\boldsymbol{k}}, \hat{\boldsymbol{p}}} \\
	=& 8e\, t_0^2\, \sin\varphi\, T \sum_{\omega_n} \left[{\sum_{\bm{k},\bm{p}}}^{\prime} \frac{1}{(\eta_{\bm{k}}^R)^2 +\omega_n^2+ \xi_{\bm{k}}^2 }\frac{1}{(\eta_{\bm{p}}^L)^2 +\omega_n^2+ \xi_{\bm{p}}^2 } \eta_{\bm{k}}^R\eta_{\bm{p}}^L \delta_{\nu\kappa}\delta_{\hat{\bm{k}},\hat{\bm{p}}} \right. \nonumber\\
	&\left. -4\omega_n^2{\sum_{\bm{k},\bm{p}}}^{\prime}\frac{\varepsilon_{\rho}^R(\bm{k})}{\{(\eta_{\bm{k}}^R)^2 +\omega_n^2+ \xi_{\bm{k}}^2 \}^2 }\frac{\varepsilon_{\lambda}^L(\bm{p})}{ \{ (\eta_{\bm{p}}^L)^2 +\omega_n^2+ \xi_{\bm{p}}^2 \}^2 } \eta_{\bm{k}}^R\eta_{\bm{p}}^L (\delta_{\lambda\rho}\delta_{\nu\kappa}-\delta_{\lambda\kappa}\delta_{\nu\rho}) \delta_{\hat{\bm{k}},\hat{\bm{p}}}
	\right]
\end{align}
As shown in the equation, there are two cases where finite Josephson current is allowed, $\nu=\kappa$, or $\lambda=\kappa$ and $\nu=\rho$ simultaneously.
The condition $\nu=\kappa$ corresponds to the selection rule $I \propto \bm{q}^L \cdot \bm{q}^R$.

For the case $\nu=\kappa$, the Josephson current is calculated as
\begin{align}
I  =& 8e\, t_0^2\, \sin\varphi\, T \sum_{\omega_n} \left[{\sum_{\bm{k},\bm{p}}}^{\prime} \frac{1}{(\eta_{\bm{k}}^R)^2 +\omega_n^2+ \xi_{\bm{k}}^2 }\frac{1}{(\eta_{\bm{p}}^L)^2 +\omega_n^2+ \xi_{\bm{p}}^2 } \eta_{\bm{k}}^R\eta_{\bm{p}}^L \delta_{\hat{\bm{k}},\hat{\bm{p}}} \right. \nonumber\\
	&\left. -4\omega_n^2{\sum_{\bm{k},\bm{p}}}^{\prime}\frac{\varepsilon_{\rho}^R(\bm{k})}{\{(\eta_{\bm{k}}^R)^2 +\omega_n^2+ \xi_{\bm{k}}^2 \}^2 }\frac{\varepsilon_{\lambda}^L(\bm{p})}{ \{ (\eta_{\bm{p}}^L)^2 +\omega_n^2+ \xi_{\bm{p}}^2 \}^2 } \eta_{\bm{k}}^R\eta_{\bm{p}}^L \delta_{\lambda\rho} \delta_{\hat{\bm{k}},\hat{\bm{p}}}
	\right] \\
	=& \frac{\pi}{eR_N}\sin\varphi \, T \sum_{\omega_n} \left[\frac{1}{\sqrt{(\eta^R)^2+\omega_n^2}}\frac{1}{\sqrt{(\eta^L)^2+\omega_n^2}} \langle \eta_{\bm{k}}^R\eta_{\bm{p}}^L \rangle_{\hat{\bm{k}}} \right. \nonumber\\
	&\left. -\omega_n^2
	\frac{1}{\{(\eta^R)^2+\omega_n^2\}^{3/2}}\frac{1}{\{(\eta^L)^2+\omega_n^2\}^{3/2}}	\langle \varepsilon_{\rho}^R(\bm{k})\varepsilon_{\lambda}^L(\bm{p})\eta_{\bm{k}}^R\eta_{\bm{p}}^L \rangle_{\hat{\bm{k}}} \delta_{\lambda\rho}
	\right]. \label{eq:apjood1}
\end{align}
If we assume $\eta^R = \eta^L =\Delta$, the Josephson current is simplified as
\begin{align}
	I=& \frac{\pi \Delta}{2e R_N}\tanh\left(\frac{\Delta}{2T}\right)\left[ \frac{\langle \eta_{\bm{k}}^R\eta_{\bm{k}}^L\rangle_{\hat{\bm{k}}}}{\Delta^2} - \frac{\langle \varepsilon_{\rho}^R(\bm{k})\varepsilon_{\lambda}^L(\bm{p})\eta_{\bm{k}}^R\eta_{\bm{p}}^L \rangle_{\hat{\bm{k}}} }{8\Delta^4}\delta_{\lambda\rho} \right] \sin\varphi .
\end{align}
The first term is equivalent to the Josephson current without the pseudospin-dependent dispersion, Eq.~(\ref{Eq:Jc_QQ_anisotropic}).
The second term represents the Josephson current carried by induced odd-frequency pairs in the two superconductors. 
It is finite for $\lambda=\rho$, where symmetries of both superconductors are identical.
Its negative sign is derived from $\mathrm{Tr}[\gamma^\nu\gamma^\lambda\gamma^\nu\gamma^\lambda]=-4$. 

For $\lambda=\kappa$ and $\nu=\rho$, the Josephson junction consists of two superconductors described by Eq.~(\ref{Eq:anisotropic_left}) for the left and Eq.~(\ref{Eq:anisotropic_right}) for the right side.
The Josephson current is calculated as
\begin{align}
	I =& \frac{\pi}{eR_N}\sin\varphi \, T \sum_{\omega_n} \left[\omega_n^2
	\frac{1}{\{(\eta^R)^2+\omega_n^2\}^{3/2}}\frac{1}{\{(\eta^L)^2+\omega_n^2\}^{3/2}}	\langle \varepsilon_{\rho}^R(\bm{k})\varepsilon_{\lambda}^L(\bm{p})\eta_{\bm{k}}^R\eta_{\bm{p}}^L \rangle_{\hat{\bm{k}}} 
	\right] \\
	=& \frac{\pi \Delta}{2e R_N}\tanh\left(\frac{\Delta}{2T}\right)\frac{\langle \varepsilon_{\rho}^R(\bm{k})\varepsilon_{\lambda}^L(\bm{p})\eta_{\bm{k}}^R\eta_{\bm{p}}^L \rangle_{\hat{\bm{k}}} }{8\Delta^4}\, \sin\varphi .
\end{align}
For the last line, we assumed $\eta^R =\eta^L =\Delta$.
The positive sign is from $\mathrm{Tr}[\gamma^\nu\, \gamma^\lambda\, \gamma^\lambda\, \gamma^\nu]=4$.

\end{widetext}

%\bibliography{paperlist}




\end{document}